\documentclass{article}
\usepackage{amsmath}

\usepackage{epsfig} 
\usepackage{subfigure}

\begin{document}

\date{}
\title{Active transport: A kinetic description based on thermodynamic grounds}
\author{S. Kjelstrup\thanks{
On leave from: Department of Chemistry, Faculty of Natural Science and
Technology, Norwegian University of Science and Technology, Trondheim,
NO-7491- Norway}, J.M. Rubi\thanks{Author to whom correspondance should be addressed. E-mail: mrubi@ub.edu}, and D. Bedeaux\small{*}, \\ 
Departament de Fisica Fonamental, Universitat de Barcelona,\\ 
Diagonal 647, 08028 Barcelona, Spain}

\maketitle
\begin{abstract}
  We show that active transport processes in biological systems can be
  understood through a local equilibrium description formulated at the
  mesoscale, the scale to describe stochastic processes. This new
  approach uses the method established by nonequilibrium
  thermodynamics to account for the irreversible processes occurring
  at this scale and provides nonlinear kinetic equations for the rates
  in terms of the driving forces.  The results show that the
  application domain of nonequilibrium thermodynamics method to
  biological systems goes beyond the linear domain.  A model for
  transport of Ca$^{2+}$ by the Ca$^{2+}$-ATPase, coupled to the
  hydrolysis of adenosine-triphosphate is analyzed in detail showing
  that it depends on the reaction Gibbs energy in a non-linear way.
  Our results unify thermodynamic and kinetic descriptions, thereby
  opening new perspectives in the study of different transport
  phenomena in biological systems. 
  
$\bf{Keywords}$: Active transport, nonequilibrium thermodynamics, kinetics, ion channels, molecular motors
\end{abstract}

\section{Introduction}
Mitchell's success in using the proton-motive force to explain energy
conversion in mitochondria (Mitchell, 1961) led to a considerable
effort over the next three decades to explore the usefulness of
nonequilibrium thermodynamics in biology
(Caplan and Essig, 1983; Westerhoff and van Dam, 1987; Walz, 1990). Even if substantial
progress was made, it was eventually concluded that the theory had
major drawbacks.  The flux-force relations that were established, were
linear, and could only describe the dynamics in a rather narrow range
near total equilibrium. Since many biological reactions have an
activation energy barrier and therefore have an intrinsically
nonlinear dynamic behavior, such an overall thermodynamic analysis
could not provide a full description of those processes.  Existing
kinetic descriptions, on the other hand, are not able to account for
the coupling between transport and chemical reaction. That coupling
has so far only rigorously been established within the framework of
Onsager's theory.  These difficulties, which already were pointed out
by Prigogine (Prigogine, 1955) and Eyring (Eyring and Eyring, 1965) in the
context of activated processes, have hampered a wider application of
nonequilibrium thermodynamics to biological systems, leaving the
complete description of active transport processes to kinetic
approaches. Kinetic descriptions take the tight coupling of transport
and chemical reaction as a starting point and can therefore not
address a varying degree of coupling or slippage.

Over the last few years, amazing new details have been revealed about
the molecular mechanism of active transport by some ATPases
(Berman, 2001).  Experimental evidence has been found that ATP
hydrolysis leads to a rotation of the central parts of the enzyme,
followed by an electric signal in the channel that spans the membrane,
attributed to the ion that is being transported, see e.g.
(Peinelt and Apell, 2004; Burzik et al., 2003; Sun et al., 2004). Active transport is
accomplished by various types of molecular motors. A theory that
relates these fluxes with driving forces in the system, however, is
still lacking (Berman, 2001).

The purpose of this paper is to show that when irreversible processes
are analyzed at a mesoscopic level, application of the method of
nonequilibrium thermodynamics leads to a complete description of the
process even in the nonlinear domain. The mesoscopic level has shorter
time and length scales than we encounter on the macroscopic level. It
is the level of description where fluctuations matter. Rate laws on
this level are stochastic of nature. We shall use mesoscopic
nonequilibrium thermodynamics (Reguera and Rubi, 2001; Vilar and Rubi,  2001) and go beyond the
linear overall or 'black box' description offered by classical
nonequilibrium thermodynamics. This gives a new perspective to the
problem of biological transport phenomena.

The paper is organized as follows. In section 2, we introduce the main features of active transport of ions and discuss the limitations of a nonequilibrium thermodynamic description of the process. In Section 3, we present a kinetic description based on thermodynamic grounds which enables us to analyze active transport processes beyond the linear domain. Finally, in Section 4 we summarize our main results. 

\section{Active transport and nonequilibrium thermodynamics}

We shall focus on the problem of active transport of an ion, the
meaning of nonlinear flux-force relations and the coupling of
vectorial and scalar forces in this context. As example we take
Ca$^{2+}$ transport in sarcoplasmic reticulum by means of the
Ca$^{2+}$-ATPase. We limit ourselves to the conversion between
chemical energy and transport of the ion, neglecting a possible
intervention of rotational energy. This does not exclude a torsional mechanism, however (Jain et al., 2004). The membrane is highly asymmetric,
about 5 nm thick, and consists of a phospholipid bilayer with proteins
embedded. Figure 1 gives a schematic picture of the Ca$^{2+}$-ATPase
embedded in the phospholipid bilayer. The reaction takes place at the
headgroup of the Ca$^{2+}$-ATPase. The ion, to be actively transported
along the protein channel, binds to a separate site and triggers the
chemical reaction, the ATP hydrolysis.  Each channel transports only a
few molecules or ions per second, and there are not many channels of
the same type in one membrane.  Experiments are done in a solution
with many vesicles or organelles, so the observation refers to the
average over the channels (Burzik et al., 2003). Thus, we have a chemical
reaction on a time scale similar to that of diffusion
(Peinelt and Apell, 2004; Burzik et al., 2003).  Can we legitimately consider this
dynamic situation with thermodynamic tools?  We shall see that the
answer is yes.

\begin{figure}[htbp]
  \centering
  \epsfig{file=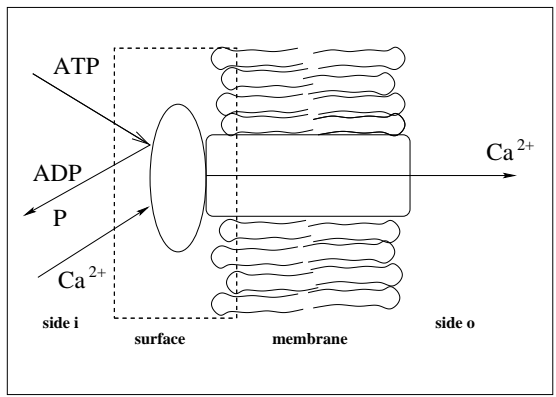, width=0.6 \textwidth}
  \caption{The Ca$^{2+}$-ATPase that transports Ca$^{2+}$ by hydrolysis of
    adenosine triphosphate (ATP) to adenosine diphosphate (ADP) and
    phosphate (P). The reaction takes place at the enzyme headgroup,
    and triggers the active transport through the protein stalk and
    channel. A surface area element is indicated by the dotted line.}
  \label{fig:1}
\end{figure}
\vspace{1 cm}

The spontaneous chemical reaction taking place at the membrane surface
in Fig.1 is, schematically:
\begin{equation}
\nonumber
\text{ATP}\left( \text{s}\right) \text{ }\rightleftharpoons \text{ADP}\left( 
\text{s}\right) \text{ + P}\left( \text{s}\right)  \label{1}
\end{equation}
where s stands for surface. The reaction Gibbs energy which arises is
\begin{equation}
\Delta G=\mu _{\text{P}}^{\text{s}}+\mu _{\text{ADP}}^{\text{s}}-\mu _{\text{%
ATP}}^{\text{s}}  \label{3}
\end{equation}
where $\mu _{j}^{\text{s}}$ is the chemical potential at the
surface of component $j$. Calcium ions are transported across the membrane from side i to side o.
The transport takes place against the chemical potential gradient as a
consequence of its coupling to the above reaction:
\begin{equation}
\nonumber
\text{Ca}^{\text{2+}}\left( \text{i}\right) \rightleftharpoons \text{Ca}%
^{2+}\left( \text{s}\right) \rightleftharpoons \text{Ca}^{2+}\left( \text{o}%
\right)  \label{2}
\end{equation}
There is probably also a countertransport of protons to keep the system's local electroneutrality (Peinelt and Apell, 2004).

Non-equilibrium thermodynamics provides a description of the
irreversible processes occurring at the surface (Albano and Bedeaux, 1987). It has
been shown that the surface behaves as an autonomous thermodynamic
system in local thermodynamic equilibrium, even in the presence of
strong driving forces (R$\o$sjorde et al., 2000).  This fact enables one to
formulate the Gibbs equation for the surface element indicated in Fig.1  (Albano and Bedeaux, 1987):
\begin{equation}
Tds^{\text{s}}= du^{\text{s}} -\mu
_{\text{ATP}}^{\text{s}}dc_{\text{ATP}}^{\text{s}}-\mu _{ 
\text{ADP}}^{\text{s}}dc_{\text{ADP}}^{\text{s}}-\mu _{\text{P}}^{\text{s}%
}dc_{\text{P}}^{\text{s}}-\mu _{\text{Ca}}^{\text{s}}dc_{\text{Ca}}^{\text{s}%
}  \label{4}
\end{equation}
Here $c^{s}_{j}$ is the surface concentration of component $j$ in
mol/mg, referring to the total unknown area of transfer per mg of
material, and $u^{\text{s}}$ is the internal energy density of the
surface. If the membrane potential contributes to transport of the
calcium ions, one has to use electrochemical potentials.  The
conservation laws for the reacting species in Eq.(\ref{1}) are
\begin{equation}
\frac{dc_{j}^{\text{s}}}{dt}= \pm r+J_{j}^{\text{i}}
\label{5}
\end{equation}
where $J_{j}^{i}$ are fluxes of species, $r$ is the reaction rate (all
in mol/s mg), and $t$ is the time. The minus sign refers to ATP and
the plus to the other species. The concentration of Ca$^{2+}$ in
mol/mg on the surface satisfies
\begin{equation}
\frac{dc_{\text{Ca}}^{\text{s}}}{dt}=J_{\text{Ca}}^{\text{i}}-J_{\text{Ca}}^{\text{o}%
}  \label{6}
\end{equation}
where $J_{Ca}^{i,o}$ are the inwards and outwards ion fluxes in
mol/s~mg. We omit the charge number in subscripts. 

For the time being, we consider an isothermal surface.  The entropy
production (in W/mg K) then follows by using Eqs.(\ref{5}) and
(\ref{6}) in combination with Gibbs equation, taking the temperature
constant and $\mu_{\text{j}}^{\text{s}}=\mu _{\text{j}}^{\text{i}}$ :
\begin{equation}
\sigma ^{\text{s}}=-r\frac{1}{T}\Delta G-J_{\text{Ca}}^{\text{o}}\frac{1}{T}%
(\mu _{\text{Ca}}^{\text{o}}-\mu _{\text{Ca}}^{\text{i}})  \label{7}
\end{equation}
The countertransport of protons to maintain the system's electroneutrality (Peinelt and Apell, 2004) can be accounted for by modifying the driven force for $Ca^{2+}$, see Eq. (20) below. From this one normally has inferred linear laws between fluxes and
driving forces (Caplan and Essig, 1983), with an expression for the flux of
Ca$^{2+}$
\begin{equation}
J_{\text{Ca}}^{\text{o}}=-\frac{L_{dr}}{T}\Delta G-\frac{L_{dd}}{T}(\mu _{%
\text{Ca}}^{\text{o}}-\mu _{\text{Ca}}^{\text{i}})  \label{8}
\end{equation}
describing linear active transport for small values of $\Delta G$ and
constant Onsager coefficients $L_{dr},L_{rr}$. Equation (\ref{7}) is
normally taken for a 'black box' description of active transport fromside i to o (Caplan and Essig, 1983). When the second term is negative, the
coupling of $J_{\text{Ca}}^{\text{o}}$ to the reaction (via $L_{dr}$)
can compensate that negative contribution thus making transport
against the concentration gradient possible.  This is the coupling
phenomenon that is referred to as active transport in the literature.
No explanation has yet been given for the nature of the coupling
(Berman, 2001), leaving the problem unsolved as how to characterize
the total process far from equilibrium.

\section{Kinetic description from thermodynamic grounds}
 
The lack of equilibrium for the chemical reaction on a time scale
similar to that for ion transport is the reason for energy
transduction between reaction and transport. This energy transduction
process shall now be described for shorter time scales at which the
system evolves from the initial to the final states. The local states
of the system correspond to successive molecular configurations and
are parametrized by an internal reaction coordinate $\gamma $ running
from 0, at which there are reactants, to 1, at which there are
products. The coordinate $\gamma$ is dimensionless. One may then
extend the assumption of local thermodynamic equilibrium to the
mesoscopic level and formulate the corresponding Gibbs equation in
$\gamma$-space:
\begin{equation}
Tds^{\text{s}}(\gamma )=-G(\gamma )dc_r(\gamma )-\mu _{\text{Ca}}^{\text{s}%
}(\gamma )dc_{\text{Ca}}^{\text{s}}(\gamma )  \label{9}
\end{equation}
where $c_r(\gamma )$ is the concentration of enzymes with a reaction
mixture characterized by coordinate $\gamma$, and
$c_{\text{Ca}}^{\text{s}}(\gamma)$ is the average concentration of
Ca$^{2+}$ over the state given by $\gamma$, both concentrations in
mol/mg. Before the reaction, the Gibbs energy becomes $G(0) \mu_{\text{ATP}}$ ; after the reaction it is $G(1) = \mu
_{\text{P}}+\mu _{\text{ADP}}$.  The relation between $\Delta G$ and
$G(\gamma)$ is shown to the left in Fig. 2.

\begin{figure}[htbp]
\centering
\mbox{
\subfigure{\epsfig{figure=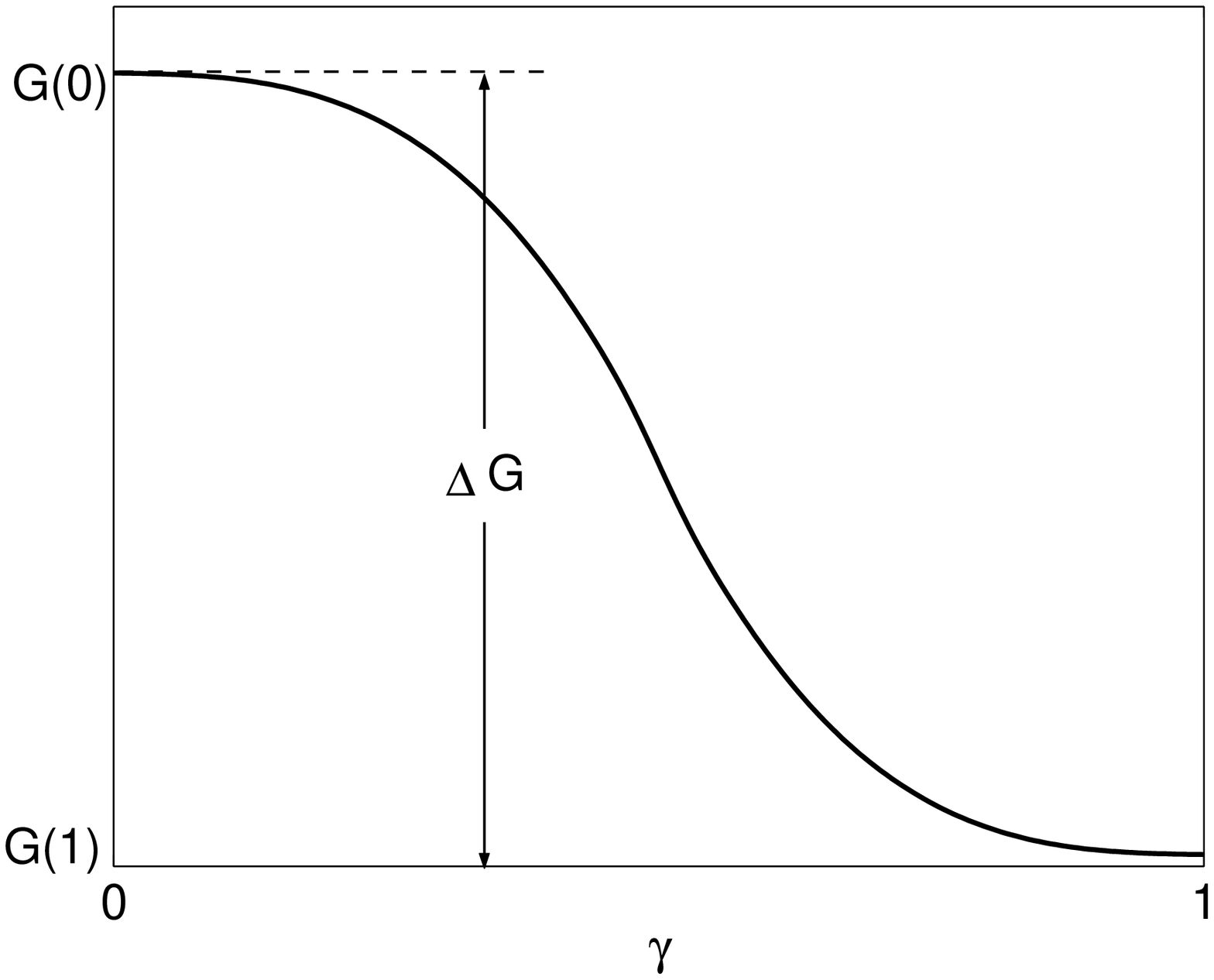,
    width=6cm, height=5cm}}
\subfigure{\epsfig{figure=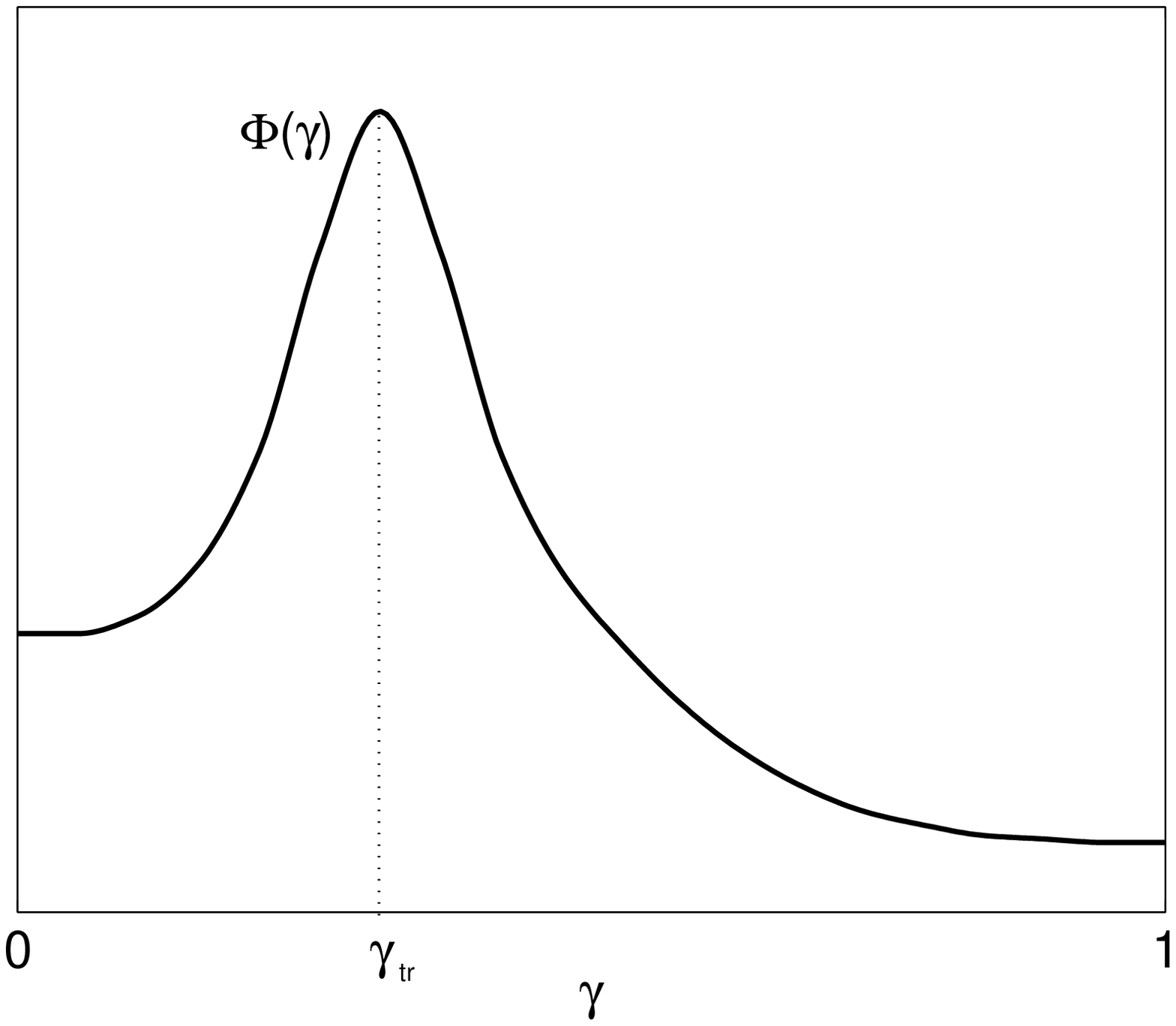,
    width=6cm, height=5cm}}}
\caption{Left, the expansion of $G(\gamma)$ in reaction coordinate space. Right, the activation energy  $\Phi(\gamma)$ with
  a peak at $\gamma_{tr}$.}
\label{fig:a_label}  
\end{figure}

Mass conservation can be expressed at any position in $\gamma$-space
through the conservation laws
\begin{equation}
\frac{dc(\gamma )}{dt}=- \frac{\partial}{\partial\gamma}r(\gamma ) \text{ \
\ and \ \ }\frac{dc_{\text{Ca}}^{\text{s}}(\gamma )}{dt}= J^{i}_{Ca}(\gamma)-J_{\text{Ca}}^{%
\text{o}}(\gamma ),  \label{10}
\end{equation}
Here $r(\gamma )$ is the local reaction flux along the reaction
coordinate.  The fluxes $J_{\text{Ca}}^{\text{i}}(\gamma )$ and
$J_{\text{Ca}}^{\text{o}}(\gamma )$ contributions to the diffusion
fluxes on sides i and o respectively, averaged over enzymes in state
$\gamma$. The integrals of the fluxes over $\gamma$ give the
corresponding global fluxes. This choice for the conservation equation
for Ca$^{2+}$ may capture the essence of the delivery of Ca$^{2+}$ to
the other side in a pump that allows slippage.

The local entropy production $\sigma ^s(\gamma)$, defined through the
entropy change per unit of time in the isothermal surface is then:
\begin{equation}
\frac{ds^s(\gamma)}{dt}=\sigma^s(\gamma)+\frac{1}{T}\frac{\partial G(\gamma)r(\gamma)}{\partial\gamma}+\frac{\mu^o_{Ca}}{T}J^o_{Ca}(\gamma)-\frac{\mu^{i}_{Ca}}{T}J^{i}_{Ca}(\gamma),
\end{equation}
then follows from Eqs.(\ref{9}) and (\ref{10})
\begin{equation}
\sigma ^{\text{s}}(\gamma )=- r(\gamma) \frac{1}{T} \frac{\partial G(\gamma )}{%
\partial \gamma }-\frac{1}{T}J^o_{\text{Ca}}(\gamma )(\mu _{\text{Ca%
}}^{\text{o}}-\mu _{\text{Ca}}^{\text{s}}(\gamma ))- \frac{1}{T}J^{i}_{\text{Ca}}(\gamma )(\mu _{\text{Ca%
}}^{\text{s}}(\gamma)-\mu _{\text{Ca}}^{\text{i}}) \label{11}
\end{equation}
As has been assumed already, there is equilibrium for the
Ca$^{2+}$-ion between the surface and the bulk solution
$\mu^{s}_{Ca}(\gamma)= \mu^{i}_{Ca}$. The local entropy production
then simplifies to
\begin{equation}
\sigma ^{\text{s}}(\gamma )=- r(\gamma) \frac{1}{T} \frac{\partial G(\gamma )}{%
\partial \gamma }-\frac{1}{T}J^o_{\text{Ca}}(\gamma )(\mu _{\text{Ca
}}^{\text{o}}-\mu _{\text{Ca}}^{\text{i}}) \label{11a}
\end{equation}
Due to the activation barrier, the reaction rate almost immediately
reaches a quasi-stationary state in which it only depends on time.
Using this property, the entropy production (4) is recovered upon
integration of Eq. (\ref{11a}) with a varying calcium flux.

We shall assume that Ca$^{2+}$ is not transported across an activation
energy barrier. This means that the ion can escape from any surface
state given by $\gamma$ with $\mu _{\text{Ca}}^{\text{s}}(\gamma)=\mu
_{\text{Ca}}^{\text{i}}$ to side o with $\mu _{\text{Ca}}^{\text{o}}$,
an explanation for why the flux depends on $\gamma$.  The local
flux-force relations that follow from Eq.(\ref{11a}), are:
\begin{eqnarray}
r(\gamma ) &=&-\frac{l_{rr}(\gamma)}{T}\frac{\partial G}{\partial \gamma }-\frac{%
l_{rd}(\gamma)}{T}(\mu _{\text{Ca}}^{\text{o}}-\mu _{\text{Ca}}^{\text{i}})
\notag \\
J_{\text{Ca}}^{\text{o}}(\gamma ) &=&-\frac{l_{dr}(\gamma)}{T}\frac{\partial G}{%
\partial \gamma }-\frac{l_{dd}(\gamma)}{T}(\mu _{\text{Ca}}^{\text{o}}-\mu _{\text{Ca%
}}^{\text{i}})  \label{12}
\end{eqnarray}
which are valid along the path going from the initial to the final
state and have been formulated under the equilibriation assumption
discussed previously. At each point of the path, the Onsager relation
$l_{rd}(\gamma)=l_{dr}(\gamma)$ is fulfilled and the coefficient do
not depend on the driving forces. The last relation will give a
non-linear equation for the flux of Ca$^{2+}$ across the membrane.

The chemical reaction is an activated process. Along the internal
coordinate, we can write the Gibbs energy as a combination of the
ideal contribution and an activation energy per mole, $\Phi (\gamma )$;
\begin{equation}
G(\gamma )=G^0+RT\ln c(\gamma )+\Phi (\gamma )  \label{13}
\end{equation}
The boundary values of $G(\gamma)$ were defined above. The standard
Gibbs energy is chosen to be $G^0=\mu_{\text{ATP}}^0$. For the
potential profile, correct boundary conditions are obtained with
$\Phi(0)=0$ and $\Phi(1)=\mu_{\text{ADP}}^0+ \mu_{\text{P}}^0 $. The
activation energy barrier with the transition state, $\gamma_{tr}$, is
illustrated in Fig.2.

We do not know the coefficients in Eqs.(\ref{12}). But, under the
condition of constant stoichiometry, or tight coupling, we can make
certain assumptions that allow us to integrate the expression for the
calcium flux.  A constant stoichiometry means that the number of
Ca$^{2+}$ transported for every reacted ATP is fixed. When the
stoichiometry is fixed, $J_{\text{Ca}}^{\text{o}}$ and $r$ are
  dependent, and the matrix of coefficients has a zero determinant.
  The number of Ca$^{2+}$ transported for every reacted ATP can also
  be defined in $\gamma$-space. We obtain:
\begin{equation}
n(\gamma )=\frac{J_{\text{Ca}}^{\text{o}}(\gamma )}{r} \frac{l_{dr}(\gamma)}{l_{rr}(\gamma)} =\frac{l_{dd}(\gamma)}{l_{rd}(\gamma)}    \label{15}
\end{equation}
In order to show how the linear, local Eqs.(\ref{12}) transform into
nonlinear global laws, we choose as thermodynamic force the gradient
of the local fugacity $z(\gamma)$,
\begin{equation}
z(\gamma)=\exp [(G(\gamma) -G^0)/RT]=c(\gamma)\exp [\Phi (\gamma )/RT]  
\label{18}
\end{equation}
The fugacity divided by the fugacity at equilibrium, can be understood
as the probability distribution function in $\gamma$-space divided by
the equilibrium distribution, $c_r(\gamma)/c_{r,eq}$.  The use of
probabilities is characteristic to a mesoscopic description.  The
local diffusion flux (\ref{12}) can then be rewritten in terms of the
new thermodynamic force
\begin{equation}
J_{\text{Ca}}^{\text{o}}(\gamma )=-\frac{nD'}{
 \exp [\Phi (\gamma )/RT]}
 \frac{\partial z(\gamma)}{\partial \gamma }- 
 \frac{l_{dd}(\gamma)}{T} 
\left[\mu_{\text{Ca}}^{\text{o}}-\mu_{\text{Ca}}^{\text{i}}%
\right]  \label{19}
\end{equation}
where we have used Eqs.(\ref{15}) and (\ref{18}) and defined
$D'=l_{rr}(\gamma)R/c(\gamma)$ in accordance with Kramers
\cite{Kramers1940}. We find the macroscopic flux in terms of the
driving forces by integrating over $\gamma$.  Using the
quasi-stationary condition for the reaction flux, and constant n, it
follows that $J_{\text{Ca}}^{\text{o}}$ is constant.  Using
furthermore that $D'$ is in good approximation constant,
we obtain:
\begin{equation}
J_{\text{Ca}}^{\text{o}}=-nD c_{\text{ATP}} \left(\exp
  \frac{\Delta G}{RT} -1 \right)-L_{
\text{Ca}}\left[ \mu_{\text{Ca}}^{\text{o}}-\mu_{\text{Ca}}^{\text{i}}\right]
\label{21}
\end{equation}
where 
\begin{equation}
D=D^{\prime }(\int_{0}^{1}\exp \Phi /RT)^{-1} \text{   and    }
L_{\text{Ca}} = \int_0^1 l_{dd}/T d\gamma 
\label{22}
\end{equation}
We have also used $\exp(G(0)-G^{0})/RT =c_{\text{ATP}}$. 

This overall diffusion flux (\ref{21}) describes active transport of
Ca$^{2+}$ for arbitrary values of $\Delta G$, and accounts thus for
the transduction process also in the nonlinear regime, which could not
previously be described through nonequilibrium thermodynamics.
Nonlinearities inherent to the kinetics of the chemical reaction
emerge in the broad description if one proceeds with a thermodynamic
analysis at the mesoscopic level and subsequently coarsens the
description by retaining only the initial and final states.  At small
values of the driving force, Eq.(\ref{21}) reduces to expression (5)
valid only in the linear regime. It is interesting to note that the
prefactor of the nonlinear term contains an Arrhenius factor
characteristic for activated processes. The first term in
Eq.(\ref{21}) is also present in the equation for active transport,
derived by the King-Altmann-Hill method (Westerhoff and van Dam, 1987).
This equation does not have a second term. The derivation and the
understanding of Eq.(\ref{21}) compared to equations that have been
used earlier to describe Ca$^{2+}$ transport, however, are very
different.

Equation (\ref{21}) can be used to plot experimental data and find the
product $nD$ and $L_{\text{Ca}}$. The value of $n$ obtained by
imposing the static head condition $J^0_{\text{Ca}}=0$ to Eq.  (17),
may be different from the one given through the relation
$n=J^o_{Ca}/r$ depending on experimental conditions. The chemical
potentials for Ca$^{2+}$ for state i or o:
\begin{eqnarray}
  \label{eq:chempot}
\mu _{\text{Ca}}^{\text{i,o}} =  \mu _{\text{Ca}}^{\text{0}} + RT \ln
c_{\text{Ca}} ^{\text{i,o}}  
\end{eqnarray}
where the ideal mixture approximation and the same standard state has
been used for both sides. The approximate difference in chemical
potentials for Ca$^{2+}$ is:
\begin{equation}
\mu _{\text{Ca}}^{\text{o}}-\mu _{\text{Ca}}^{\text{i}}=RT\ln \frac{%
c_{\text{Ca}}^{\text{o}}}{c_{\text{Ca}}^{\text{i}}}\simeq RT\frac{%
c_{\text{Ca}}^{\text{o}}-c_{\text{Ca}}^{\text{i}}}{c_{\text{Ca}}^{
\text{i}}}  \label{14}
\end{equation}
This driving force can be modified to also account for the
countertransport of protons. The effective driving force for exchange
of Ca$^{2+}$ and H$^{+}$ becomes
 \begin{equation}
\Delta \mu =RT\ln \frac{
c_{\text{Ca}}^{\text{o} }c_{\text{H}}^{\text{i} }}{c_{\text{Ca}}^{\text{i}}c_{\text{H}}^{\text{o} }}  \label{21}
\end{equation}
Only a zero difference in pH between the sides has no influence on this force. 

In the derivation of the flux equations (\ref{12}), we have made the
assumption that the surface is isothermal. Otherwise the model is
general in the way that it allows slippage. We have further assumed
that Ca$^{2+}$ transport across the membrane is not an activated
process, but rather that its contribution to the total flux is simply
linear in its corresponding driving force, with a transport
coefficient independent from the activation energy. Therefore, the
only nonlinearity comes from the kinetics of the chemical reaction.
This assumption may be valid for the concentrations of calcium that
are physiologically relevant. 

In order to integrate the flux on the mesoscopic level, we introduced
one more assumption, namely that of a tight pump. This assumption is
thus a requirement for Eq.(\ref{21}). The assumption may
hold for an isothermal surface, but there are many conditions for
which this is not true (Berman, 2001).

The theory we have presented could be useful in describing other types
of active transport in biological systems, as the ones taking place in
molecular motors for which translational (Bedeaux et al., 2004) or
rotational motion may be induced at the expense of the energy provided
by the chemical reaction (Sun et al., 2004). It can also be used for a
more detailed look at pump slippage, believed to occur for high
driving forces, and indicated by a variable stoichiometry
(Zoratti et al., 1986; Walz, 1990). One advantage of the present formulation,
not contained in kinetic approaches, is that the energy dissipated as
heat (Berman; 2001) can be directly calculated from
$T\sigma^{\text{s}}$. A large value of $T\sigma^{\text{s}}$ suggests
that the surface may be nonisothermal, and $T\sigma^{\text{s}}$ has a
meaning also for nonlinear flux-force relations.

\section{Conclusions}

The main aim of this paper has been to show that active transport
processes can be described by means of mesoscopic nonequilibrium
thermodynamics (Reguera and Rubi, 2001; Vilar and Rubi, 2001). We have been able to  describe the
coupling between the chemical reaction and transport, 
which is mandatory in a complete thermodynamic description of this
type of system, and which was previously not obtained outside the
range of linear flux-force-relations (Berman, 2001).

We have shown that when local
equilibrium is formulated at shorter time scales, the overall
diffusion flux is nonlinear in the macroscopic driving force. A new
way to plot experimental data has thus been given, applicable for
large values of $\Delta G$. The formulation we propose is of a
mesoscopic nature and may incorporate the presence of fluctuations in
the dynamics consistent with a Fokker-Planck description (Vilar and Rubi, 2001).
We have somewhat arbitrarily taken active transport of Ca$^{2+}$ as an
example, but the scheme proposed could be transferable to many kinds
of transport processes that obtain their energy from chemical
reactions such as molecular motors. This shows that nonequilibrium
thermodynamic methods are not restricted to the linear response domain
(J$\ddot u$licher et al., 1997) but can be used in a broader context to perform a
description of the dyamics also in the nonlinear domain. We hope in
this way to provide a general scenario in which nonlinear transport
processes occurring in biological system can be dealt with.

\subsubsection*{Acknowledgment}

S.K. is grateful to MECD Grant number SAB 2002-0191 that made possible
her sabbatical stay in University of Barcelona. D.B. is grateful to
MECD Grant number SAB 2003-0051. This work was partially supported by
DGICYT of the Spanish Government under Grant No. PB2002-01267.

\noindent ALBANO, A.M. \& BEDEAUX, D. (1987). Non-equilibrium electro-thermodynamics   of polarizable multicomponent fluids with an interface. \textit{Physica A}
  \textbf{147}, 407-435.\\ 
  
\noindent BEDEAUX, D., KJELSTRUP, S.\& RUBI, J.M. (2004). Nonequilibrium translational effects in evaporation and condensation.
  \textit{J. Chem. Phys.} \textbf{119}, 9163-9170. \\
  
\noindent BERMAN, M.C. (2001). Slippage and uncoupling in P-type cation pumps; implications for energy transduction mechanisms and regulation of metabolism. \textit{Biochim. Biophys.
    Acta} \textbf{1513}, 95-121.\\
    
\noindent BURZIK, C., KAIM, G., DIMROTH, P., BAMBERG, E. \&
  FENDLER, K. (2003). Charge displacements during ATP-hydrolysis and synthesis of the Na$^{+}$-transporting $F_{0}F_{1}$-ATPase of llyobacter tartaricus. \textit{Biophysical Journal}, \textbf{85}, 2044-2054.\\
  
\noindent CAPLAN, S.R. \& ESSIG, A. (1983) \textit{Bioenergetics
and linear nonequilibrium thermodynamics. The steady state}. (Harvard
University Press, Cambridge, Massachusetts).\\

\noindent EYRING, H. \& EYRING, E. (1965) \textit{Modern Chemical
    Kinetics.} (Chapman and Hall, London).\\
    
\noindent JAIN, S., MURUGAVEL, R. \& HANSEN, L.D. (2004). ATP synthase and the torsional mechanism: Resolving a 50-year-old mystery.
 \textit{Curr. Sci.} \textbf{87}, 16-19.\\

\noindent J$\ddot U$LICHER, F., ADJARI, A. \& PROST, J. (1997). Modeling molecular motors.
  \textit{Rev. Mod. Phys} \textbf{69}, 1269-1281.\\

\noindent KRAMERS, H.A. (1940). Brownian motion in a field of force and the diffusion model of chemical reactions  \textit{Physica}
  \textbf{7}, 284-304.\\
  
\noindent MITCHELL, P. (1961). Coupling of phosphorylation to electron and hydrogen transfer by a chemi-osmotic type mechanism. \textit{Nature (London)}, 
\textbf{191}, 144-148.\\

\noindent PEINELT, C. \& APELL, H.J. (2004). Time-resolved charge movements in the sarcoplasmatic reticulum Ca-ATPase. \textit{Biophysical
    J.} \textbf{86}, 815-824.\\
 
\noindent PRIGOGINE, I. (1965) \textit{Thermodynamics of
    Irreversible Processes.} (Charles C. Thomas, Springfield).\\
    
\noindent REGUERA, D. \& RUBI, J.M. (2001). Kinetic equations for diffusion in the presence of entropic barriers. \textit{Phys. Rev.
    E} \textbf{64}, 061106 (1-8).\\
    
\noindent R\O SJORDE, A., FOSSMO, D.W., BEDEAUX, D., KJELSTRUP, S. \&
  HAFSKJOLD, B. (2000). Nonequilibrium Molecular Dynamics Simulations of Steady-State Heat and Mass Transport in Condensation: I. Local Equilibrium.   \textit{J. Colloid Interf. Sci.} \textbf{232},
  178-185.\\
  
\noindent SUN, S.X., WANG, H. \& OSTER, G. (2004). Asymmetry in the $F_{1}-ATPase$ and Its Implications for the Rotational Cycle.
  \textit{Biophysical J.} \textbf{86}, 1373-1384.\\
  
\noindent VILAR, J.M. \& RUBI, J.M. (2001). Thermodynamics beyond local equilibrium. \textit{Proc. Natl.
    Acad. Sci. USA} \textbf{98}, 11081-11084.\\
    
\noindent WALZ, D. (1990). Biothermokinetics of processes and energy conversion. \textit{Biochim. Biophys. Acta}
  \textbf{1019} 171-224.\\  
  
\noindent  WESTERHOFF, H.V. \& VAN DAM, K. (1987) \textit{
Thermodynamics and control of biological free-energy transduction}
(Elsevier, Amsterdam).\\

\noindent ZORATTI, M., FAVARON, M., PIETROBON, D. \&
   AZZONE, G.F. (1986). Intrinsic uncoupling of mitochondrial proton pumps.  \textit{Biochemistry} \textbf{25}, 760-767.\\

\end{document}